\documentclass[12pt]{article}

\usepackage{graphicx}
\usepackage{amsmath}

\title{Intensity Coding in Two-Dimensional Excitable Neural Networks}

\author{Mauro Copelli\\ 
Laborat\'orio de F\'{\i}sica Te\'orica e Computacional \\
Departamento de F\'{\i}sica, Universidade Federal de Pernambuco \\
50670-901, Recife, PE, Brazil
\and Osame Kinouchi \\
Laborat\'orio de Sistemas Neurais \\
Departamento de F\'{\i}sica e Matem\'atica,\\
Faculdade de Filosofia, Ci\^encias e Letras de Ribeir\~ao Preto \\
Universidade de S\~ao Paulo, 14040-901, Ribeir\~ao Preto, SP, Brazil
}

\begin{document}

\maketitle

\abstract{In the light of recent experimental findings that gap
junctions are essential for low level intensity detection in the
sensory periphery, the Greenberg-Hastings cellular automaton is
employed to model the response of a two-dimensional sensory network to
external stimuli. We show that excitable elements (sensory neurons)
that have a small dynamical range are shown to give rise to a collective
large dynamical range. Therefore the network transfer (gain) function
(which is Hill or Stevens law-like) is an emergent property generated
from a pool of small dynamical range cells, providing a basis for a
``neural psychophysics''. The growth of the dynamical range with the
system size is approximately logarithmic, suggesting a functional role
for electrical coupling. For a fixed number of neurons, the dynamical
range displays a maximum as a function of the refractory period, which
suggests experimental tests for the model. A biological application to
ephaptic interactions in olfactory nerve fascicles is proposed.}

\section{Introduction}
\label{intro}

Experimental results show that individual receptor neurons have a
small dynamical range~\cite{Firestein93,Rospars00}. Given a properly
defined stimulus intensity $r$ (e.g. light intensity for rods in the
retina, or odorant concentration for sensory neurons in the olfactory
epithelium), the response $f(r)$ of a single neuron (e.g. the spike
frequency) saturates at a value $f_{max}$ for sufficiently large
values of $r$. Defining $r_{10}$ and $r_{90}$ as the stimulus values
for which the response attain $10\%$ and $90\%$ of $f_{max}$, the
dynamical range $\Delta r$ is usually defined~\cite{Firestein93} as
\begin{equation}
\label{dynrange}
\Delta r = \log_{10}\left(\frac{r_{90}}{r_{10}}\right),
\end{equation}
in log units. In other words, the dynamical range roughly measures the
number of decades over which the stimulus can be properly
discriminated\footnote{An alternative definition employs $5\%$ and
$95\%$ of the saturation value for evaluating $\Delta r$. This is of
course just a matter of choice. Our results remain qualitatively the
same in either case.}, discarding stimuli too faint to be considered
``detected'' ($r<r_{10}$) or too close to saturation ($r>r_{90}$). In
olfactory receptor cells of the tiger salamander~\cite{Firestein93}
and the frog~\cite{Rospars00}, $\Delta r \sim 1$~log unit (10~dB).

The psychophysics literature, on the other hand, shows that our
ability to discriminate external signals of different kinds covers a
{\em large\/} dynamical range~\cite{Stevens}. This is reflected in
phenomenological psychophysical laws used to fit data, which state
that the psychological perception of a given stimulus $r$ increases as
$\sim r^\alpha$, $\alpha<1$ (Stevens' law), as $\sim \log(r)$
(Weber-Fechner law), or as $\sim r^\alpha/(c^\alpha + r^\alpha)$ (Hill
function, which may be thought of as a saturating Stevens' law). That
is, differently from individual sensory neurons, organisms perform
{\em signal compression\/}, being able to process environmental
stimuli whose intensities usually span several orders of magnitude.

How can we reconcile these results? How could this signal compression
be physically implemented considering that, at the very first stage of
the signal processing, individual sensory neurons have a small
dynamical range?

Following a conjecture put forward by Stevens~\cite{Stevens}, we
propose that this signal compression occurs already at the periphery
of the nervous system, that is, at the first ``relevant'' synaptic
level. In our model, this processing comes out naturally as a result
of the collective phenomenon of {\em non-linear amplification\/} in
excitable networks. While {\em individual\/} sensory neurons have a
small dynamical range, an {\em array\/} of connected neurons can
collectively exhibit a very large dynamical range.

The motivation for the model comes from recent experimental findings
showing that electrical synapses by gap junctions are present in the
periphery of different sensory systems, such as the olfactory
glomeruli~\cite{Kosaka02,Zhang03} and the retina~\cite{Deans02}. With
this information at hand, we model a given sensory periphery as a
two-dimensional array of excitable elements which are laterally
connected by electrical synapses. Experimentally, the topology of the
neural network connected by gap junctions is still not known, and
specific details probably depend on the particular sensory system
under consideration. So the two-dimensional lattice can be considered
as the natural first step towards a biologically acceptable model. As
a possible biological application, it could represent the cross
section of a nerve fascicle containing highly packed unmyelinated
axons coupled by electric interactions, which target a common
secondary neuron, as found in the olfactory periphery.  The
one-dimensional case, which may be thought of as an approximation for
networks where neurons are coupled to two neighbors in average, has
been previously addressed in Ref.~\cite{Copelli02}.

In section~\ref{model} we introduce the cellular automaton model used
for each excitable element and briefly review results from the
literature. Our results are discussed in section~\ref{results}, while
section~\ref{conclusion} brings our concluding remarks and suggestions
for experimental tests.

\section{The model}
\label{model}

In order to simulate large networks, we make use of the simplest
possible model for an excitable element: an $n$-state cellular
automaton (CA). The state of the $i$-th cell ($i=1,\dots,N$) at
discrete time $t$ is denoted by $x_i(t) \in \{0,1,\ldots,n-1\}$. A
field $h_i(t)\in \{0,1 \}$ indicates whether the stimulus at site $i$
is supra- or sub-threshold.  The CA model contains two ingredients:
\begin{enumerate}
\item a cell spikes in the next time step ($x_i(t+1)=1$) only if it is
currently stimulated ($h_i(t)=1$) AND in its resting state
($x_i(t)=0$)
\item after a spike, a refractory period takes place ($x_i =
2,3,\ldots,n-1$), during which no further spikes occur, until the cell
returns to its resting state (formally, if $x_i(t)>0$, then $x_i(t+1)=
[x_i(t)+1]\mod n$ with probability one)
\end{enumerate}

This corresponds to one of the several variants of the model proposed
by Greenberg and Hastings~\cite{Greenberg78}. It was initially
designed to provide a simple explanation for the mechanisms underlying
pattern formation in excitable media. Indeed, the onset of
self-sustained or re-entrant activity (with e.g. spiral waves) in a
plethora of systems (including from heart tissue to chemical
reactions) has been a research topic of considerable interest (see
e.g. Refs.~\cite{Kaplan88,Bardou96} and references therein).
Different flavors of the Greenberg-Hastings cellular automaton (GHCA)
have been used in a variety of applications: collective oscillations
of pyramidal cells in the hippocampus~\cite{Traub99b,lewis00};
noise-induced memory~\cite{Chialvo} and the evolution of the HIV
infection~\cite{Zorzenon00}. In this paper we address an apparently
unaccounted feature of the widely studied GHCA: the collective
enhancement of the dynamical range of the single cells.

We model the response of a sensory neural network to some external
input, and assume that this response vanishes if input is removed.
This is accomplished with the rules defined above which, together with
a resting initial condition ($x_i(t=0)=0$, $\forall i$) and the
synchronous update of the lattice, prevent self-sustained activity
from occurring~\cite{lewis00}. We also assume that the input signal
$I_i(t)$ (which may have been preprocessed by earlier layers in a
biological network) arriving on cell $i$ at time $t$ is a Poisson
process of supra-threshold events of stereotyped unit amplitude. That
is, $I_i(t)= \sum_n \delta\left( t,t^{(i)}_n\right)$ where
$\delta(a,b)$ is the Kronecker delta and the time intervals
$t^{(i)}_{n+1}-t^{(i)}_{n}$ are distributed exponentially with average
(input rate) $r$, measured in events per second. At each time step,
the probability of arrival (per neuron) of an external input is
therefore

\begin{equation}
\label{lambdar}
\lambda(r)=1-e^{-r\tau}\; , 
\end{equation}
where $\tau=1$~ms coincides with the approximate duration of a spike
and is the time scale adopted for a time step of the model. With this
choice, the number of states $n$ of the GHCA corresponds roughly to
the refractory period (measured in ms) of the sensory neurons (or the
sensory axons).

The network connectivity is implicitly defined by the field $h_i$. For
uncoupled cells, we have simply $h_i^{(u)}(t) = I_i(t)$, i.e. cells
are excited only by external stimuli. We model the coupling via gap
junctions with the field
\begin{equation}
\label{field}
h_i^{(c)}(t) = 1-\left[1-I_i(t) \right]
\prod_{j} \left[1-\delta\left( x_j(t),1\right) \right]\; ,
\end{equation}
i.e. cells are excited by external stimuli or at least one spiking
cell $j$ (where $j$ runs over the neighborhood of $i$). We have
employed periodic as well as open boundary conditions, with
indistinguishable results. The number of sites in the regular lattice
with coordination number $z$ is $N=N_x N_y$.

The response $F(r)$ of the lattice to a given stimulus intensity $r$
is given by the mean firing rate per cell in a long time interval $T$
(typically we have used $T={\cal O}(10^5)$ time steps). This should be
compared to the mean firing rate per cell $f$ of an ensemble of
uncoupled cells, whose dependence on $r$ can be exactly calculated:
\begin{equation}
\label{f}
f(r) = \frac{\lambda(r)}{1+(n-1)\lambda(r)}\; .
\end{equation}
This linear saturating function will be the benchmark against which
results from the lattices should be compared. Note that, with the
definition of Eq.~\ref{dynrange}, uncoupled cells have a dynamical
range
\begin{equation}
\label{dynrangesingle}
\Delta r= \log_{10}
\left\{
\frac{\ln\left[ 1+\frac{9}{n} \right]}{\ln\left[ 1+\frac{1}{9n} \right]}
\right\} \; ,
\end{equation}
which is a smooth curve and, for moderate values of $n$, is well
approximated by the asymptotic value $\Delta r
\stackrel{n\to\infty}{\simeq} \log_{10}(81)\simeq 19$ dB. This
dynamical range is slightly larger than the experimental results, but
still too small to account for psychophysical data~\cite{Stevens}.

\section{Results and Discussion}
\label{results}

\begin{figure}[!tb]
\begin{center}
\hspace*{-0.5cm}
\includegraphics[width=0.65\textwidth,angle=-90]{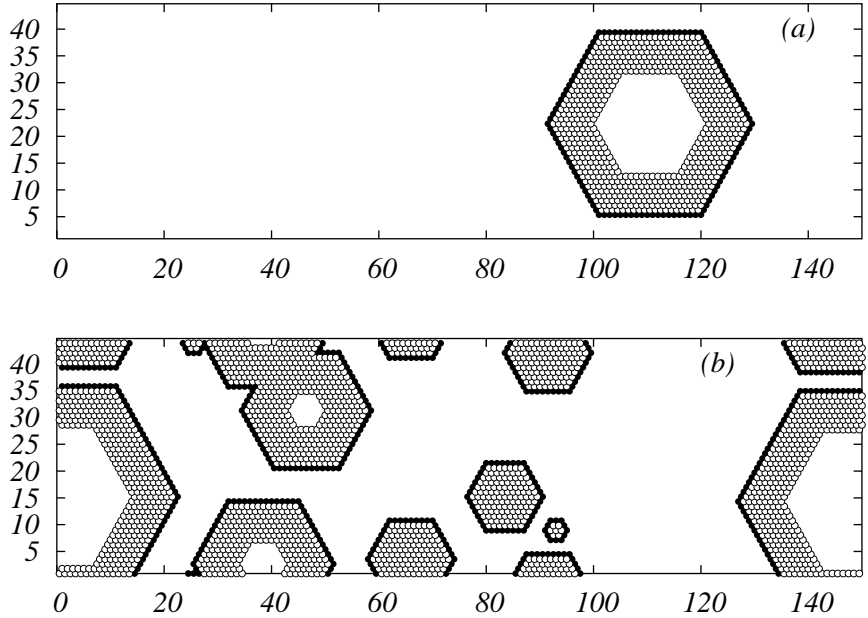}
\caption{\label{fig:snapshot}Snapshots of a triangular $150\times 50$
lattice where $10-$state CA interact with their $z=6$ nearest
neighbors (periodic boundary conditions). Closed circles: $x_i(t)=1$
(spiking). Open circles: $2<x_i(t)<n$ (refractory). Background:
$x_i(t)=0$ (resting). Panel (a): $r=0.005$~s$^{-1}$. Panel (b):
$r=0.1$~s$^{-1}$.}
\end{center}
\end{figure}

For very low values of $r$, the stimulation of a single site generates
a wave of activation which propagates through the entire lattice (see
Fig.~\ref{fig:snapshot}(a)). Before another stimulus is produced, the
wave front is annihilated at the walls (with open boundary conditions)
or when it touches itself (with periodic boundary conditions), because
sites which have spiked in the last $n-1$ time steps cannot spike
again due to the refractory period. In this regime, a single stimulus
produces $N$ spikes, yielding an $N$-fold amplification,
\begin{equation}
\label{Frto0}
F \stackrel{r\simeq 0}{\simeq} N\lambda \simeq Nr\tau \; .
\end{equation}

At the other extreme, for very high values of $r$ the propagation of
waves is suppressed by excess of external stimulus: in this regime,
the coupling becomes irrelevant since each cell is already spiking at
its maximal rate, 
\begin{equation}
\label{Frtoinfty}
F \stackrel{r\to\infty}{\simeq} f \simeq \frac{1}{n} \equiv F_{max} \; ,
\end{equation}
yielding no amplification, $F/f \simeq 1$.

In between those two regimes, waves are created, propagate and
annihilate when meeting one another, yielding complex geometric
patterns which appear from the remains of partially destroyed waves
(see Fig.~\ref{fig:snapshot}(b)). The interplay between wave creation
and wave annihilation provides the self-limited mechanism by which the
amplification $F/f$ decreases smoothly from ${\cal O}(N)$ (for small
$r$) to ${\cal O}(1)$ (for large $r$). This is a mechanism which leads
to signal compression.

\begin{figure}[!bt]
\begin{center}
\includegraphics[width=0.65\textwidth,angle=0]{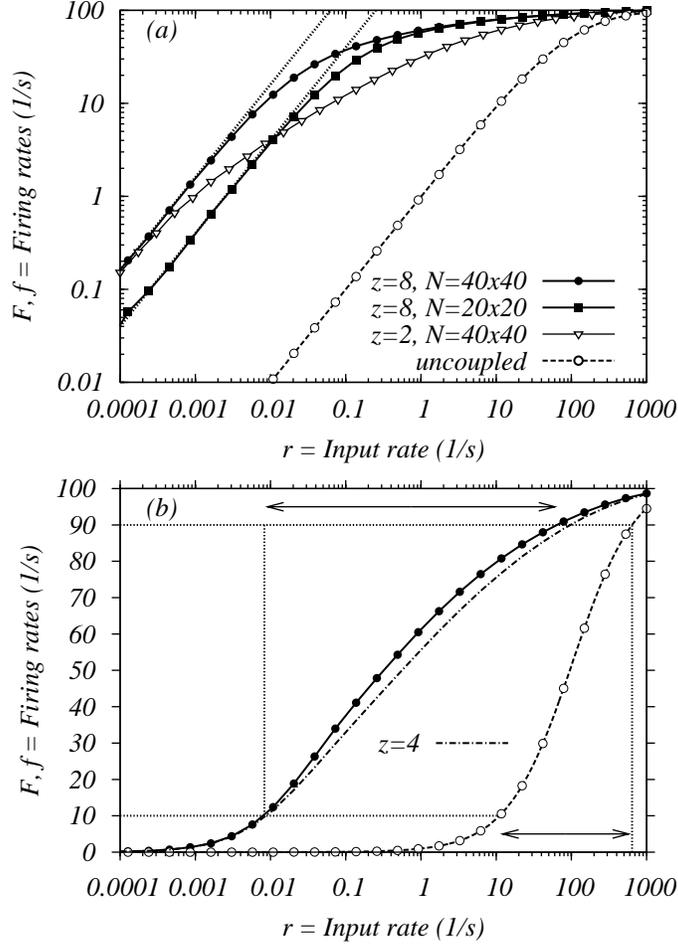}
\caption{\label{fig:Fr}Response curves for square lattices of
$10-$state automata. (a) coupled ($F$, filled symbols) and uncoupled
($f$, open symbols) cases for $z=8$ neighbors. The dashed line
corresponds to the exact solution (Eq.~\ref{f}), while the dotted
lines are the linear approximation, Eq.~\ref{Frto0}. Open triangles
are results for a one-dimensional coupled system; (b) Linear-log
version of (a), with the dot-dashed line representing a square lattice
with $N=40\times 40$ and $z=4$ neighbors. Arrows indicate the
dynamical range.}
\end{center}
\end{figure}

Fig.~\ref{fig:Fr}(a) shows the result of numerical simulations for
square lattices with $z=8$ and $n=10$ in a log-log plot. The linear
response (Eq.~\ref{Frto0}) for low values of $r$ can be clearly seen
at the leftmost portion of the graph, requiring lower values of $r$
for larger system sizes (for a fixed low value of $r$, the larger the
system the more probable it is for two excitable waves to
interact). As $r$ increases, wave annihilation starts competing with
wave creation, bending down the nonlinear response. Finally,
saturation ensues for values of $r$ close to $1/n$, making the coupled
and uncoupled cases nearly indistinguishable. The enhancement of the
dynamical range becomes clear when we plot $F(r)$ in a linear-log plot
[Fig.~\ref{fig:Fr}(b)]: since the amplification due to coupling is
larger for lower $r$, the curves stretch upward, compressing more
decibels of stimuli in the same $[0.1F_{max},0.9F_{max}]$ response
interval.

The dimensionality of the lattice has a significant impact on the
collective response, as exemplified by the results for a
one-dimensional chain in Fig.~\ref{fig:Fr}. The differences can be
understood as consequences of the following factors: a) in 1D wave
fronts have a constant size (2 sites), while in 2D their size grows
with time; b) because of a), waves in 2D can be partially destroyed
(as exemplified in Fig. 1), while in 1D they are completely
annihilated upon collision; c) the time it takes for a wave to reach
the borders scales with $N$ in 1D, but with $\sqrt{N}$ in 2D. The
combination of factors (a), (b) and (c) gives rise to finite size
effects which are much stronger in 2D than in 1D. As Fig.~\ref{fig:Fr}
shows, for example, 1D chains with 1600 or 400 neurons would have
essentially the same dynamical range, since finite size effects appear
below $r_{10}=10$~Hz. For 2D lattices, on the other hand, the
dynamical range changes. 

We have simulated two-dimensional square lattices with $z=4$ and $z=8$
neighbors, as well as triangular lattices with $z=6$. For a fixed
dimensionality, the differences in coordination number are not
important to our investigations, as can be seen in Fig.~\ref{fig:Fr}
(the curve for $z=6$ falls between those of $z=4$ and $z=8$). In what
follows, we focus on results for $z=8$.

\begin{figure}[!tb]
\begin{center}
\hspace*{-0.5cm}
\includegraphics[width=0.6\textwidth,angle=-90]{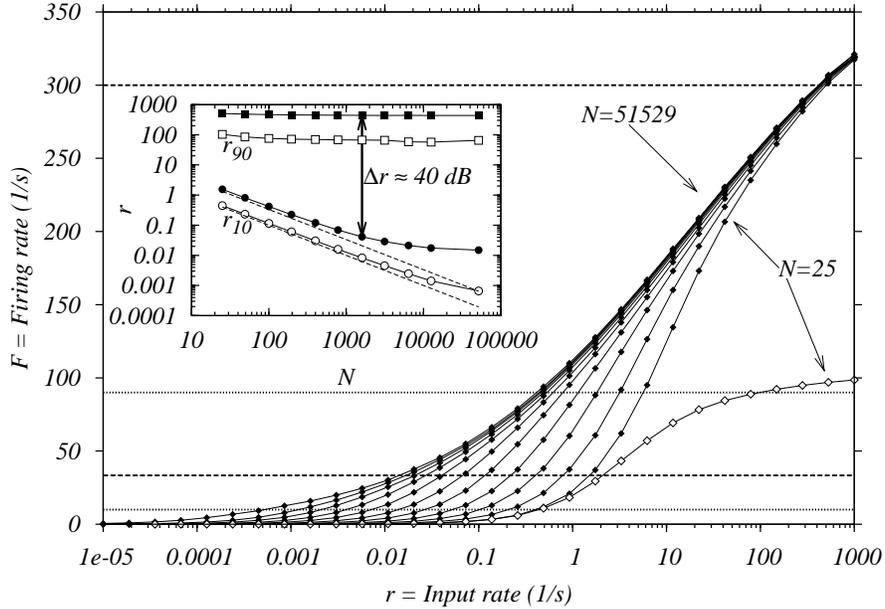}
\caption{\label{fig:several}Response curves of $3-$state (filled
diamonds) and $10-$state (open diamonds) automata for different
lattice sizes $N$ ($N=25$ for $n=3$ and 10, up to $N=51529$ for
$n=3$). The upper (lower) horizontal line corresponds to $90\%$
($10\%$) of the saturation firing rate for $n=3$
($F_{max}=1/3$~ms$^{-1} \simeq 333$~s$^{-1}$, thick dashed) and $n=10$
($F_{max}=1/10$~ms$^{-1} \simeq 100$~s$^{-1}$, thin dotted). Inset:
sensitivity ($r_{10}$, circles) and saturation ($r_{90}$, squares)
levels (filled symbols for $n=3$, open symbols for $n=10$) as a
function of the number of sites $N$ in the lattice. The vertical arrow
illustrates the dynamical range for a $40\times 40$ lattice with
$n=3$. Dashed lines correspond to Eq.~\ref{r10aprox}.}
\end{center}
\end{figure}

For a given refractory period $n$, the size of the lattice regulates
the crossover from the linear to the nonlinear response. If the
sensitivity level $r_{10}$ lies within the linear regime (as it does,
for instance, in Fig.~\ref{fig:Fr}), then the dynamical range will be
affected by the system size. It is important to note that this is not
an artificial feature due to the somewhat arbitrary definition of
$\Delta r$. Any reasonable definition of the dynamical range will
reveal a sensitivity gain (decrease in $r_{10}$) for increasing values
of $N$, as depicted in the family of curves of
Fig.~\ref{fig:several}. Note that the saturation level $r_{90}$ also
changes, but much less than $r_{10}$, leading to an enhancement of the
dynamical range.

\begin{figure}[!tb]
\begin{center}
\hspace*{-0.5cm}
\includegraphics[width=0.6\textwidth,angle=-90]{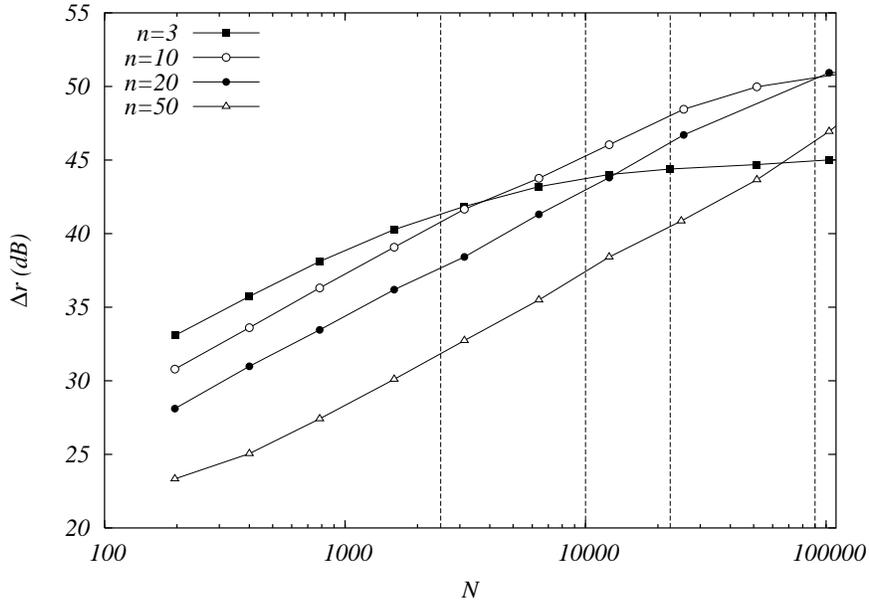}
\caption{\label{fig:dynrange}Dynamical range as a function of the
  lattice size for different values of $n$. The vertical lines
  indicate values of $N$ to be considered in more detail
  subsequently.}
\end{center}
\end{figure}

\begin{figure}[!tb]
\begin{center}
\hspace*{-0.5cm}
\includegraphics[width=0.6\textwidth,angle=-90]{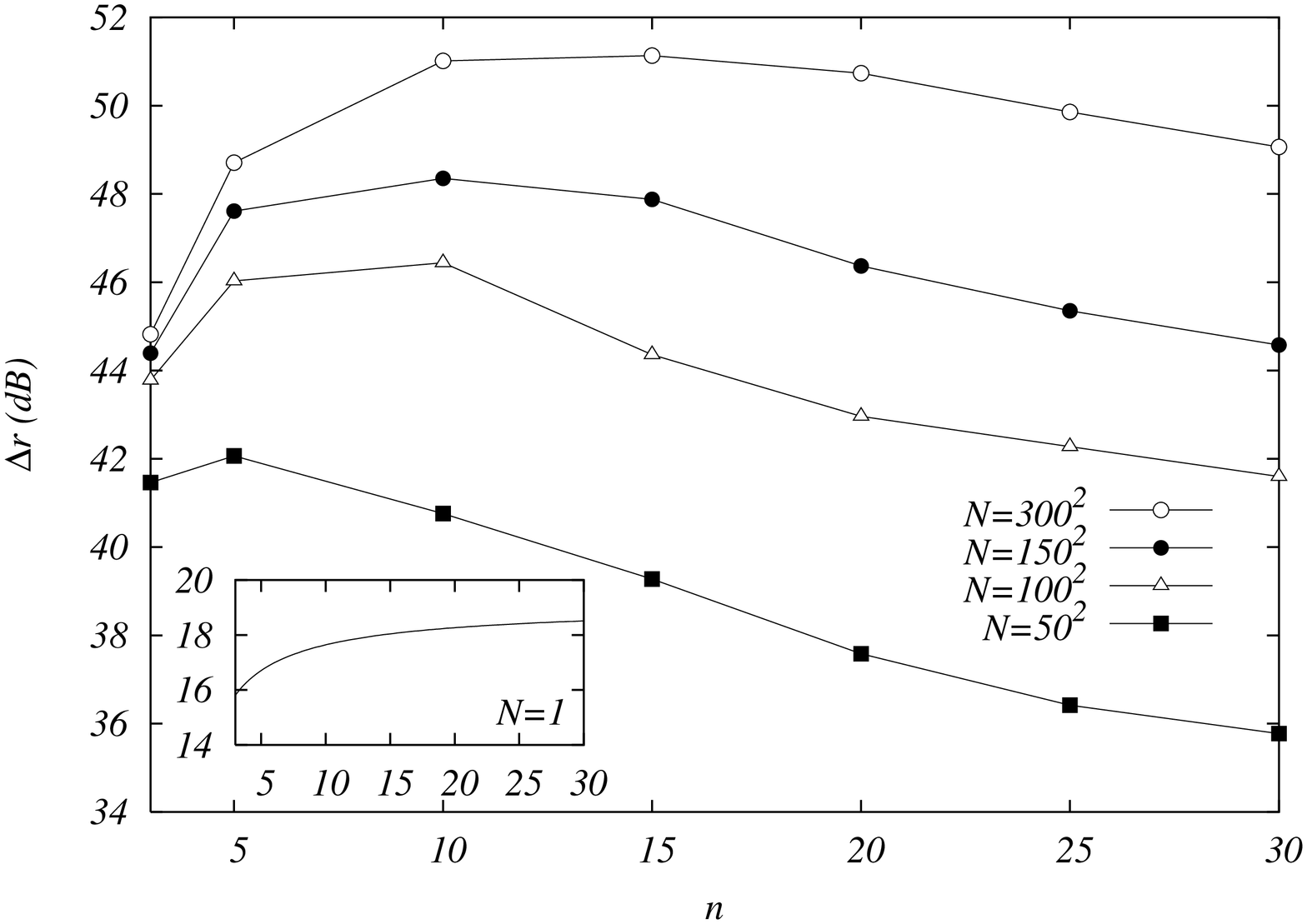}
\caption{\label{fig:dynrangenstate}Dynamical range as a function of
  the refractory period $n$ for different values of the lattice
  size. Inset: exact expression for an isolated neuron
  (Eq.~\ref{dynrangesingle}).}
\end{center}
\end{figure}

It is also worth mentioning that, since $r_{10}$ and $r_{90}$ are
defined only relative to a saturation response, these quantities will
depend on $n$. Fig.~\ref{fig:several} and its inset illustrate this
fact for $n=3$ (filled symbols) and $n=10$ (open symbols). Note that
the relative meaning of ``sensitivity'' and ``saturation'' explains
the following apparent contradiction: even though the $n=3$ curves lie
{\it above\/} the $n=10$ curves (for the same system size), their
sensitivity and saturation levels {\it also lie above\/} those of the
$n=10$ case. In other words, if $r_{10}$ is assumed to lie in the
linear regime, then Eqs.~\ref{lambdar} and \ref{Frto0} yield a
decreasing dependence on $N$ {\it and\/} $n$:
\begin{equation}
\label{r10aprox}
r_{10} \simeq
- \frac{1}{\tau}\ln\left(1-\frac{0.1}{nN}\right)\; .
\end{equation}
This approximation is acceptable for moderate values of $N$ (above
which the interference among excitable waves invalidates the linearity
assumption) and improves for increasing $n$ (since this lowers
$r_{10}$ into the linear region --- see inset of
Fig.~\ref{fig:several}).

The above discussion only emphasizes that the dynamical range stands
out as a better quantity to look at than $r_{10}$ and $r_{90}$, being
a bona fide {\it dimensionless\/} variable which could be compared to
experimental data. The dependence of $\Delta r$ on the system size can
be seen in Fig.~\ref{fig:dynrange}. Note that an expansion in
Eq.~\ref{r10aprox} for $N\gg 0.1/n$ yields 

\begin{equation}
\label{loggrowth}
\Delta r (N)\simeq \log_{10}(r_{90}) + \log_{10}(10\tau n N)\; .
\end{equation}
If we neglect the (weak) dependence of $r_{90}$ on $N$, this
expression explains the approximately logarithmic growth in
Fig.~\ref{fig:dynrange} before the curves level off. For larger values
of $N$, the linearity assumption for $r_{10}$ breaks down and
Eqs.~\ref{r10aprox} and~\ref{loggrowth} are no longer valid.

In Fig.~\ref{fig:dynrange} one observes that for low values of $N$
(say, up to ${\cal O}(10^3)$), lower values of $n$ imply larger
dynamical ranges (this cannot be inferred from Eq.~\ref{loggrowth}
since it omits the unknown and strong dependence of $r_{90}$ on
$n$). As $N$ grows, however, the curves in Fig.~\ref{fig:dynrange}
start leveling off for increasing values of $n$. This simple fact has
very interesting consequences.  Suppose that we fix the system size at
one of the four values specified by the vertical lines in
Fig.~\ref{fig:dynrange}. If we now vary the refractory period $n$, we
obtain the four curves displayed in Fig.~\ref{fig:dynrangenstate},
which have a maximum.

If we accept the reasoning that a large dynamical range is a desirable
property for a live organism, then the results presented so far
suggest two biological outcomes: first, that it would be advantageous
for the organism to have as many sensory cells as possible.  By
electrically coupling them, an increase of more than 100\% in
dynamical range could be obtained, as compared to isolated neurons
(see inset of Fig.~\ref{fig:dynrangenstate}). This is supported by the
experimental evidence that gap junctions are present in the sensory
periphery (which was in fact what motivated the
model~\cite{Copelli02}). Recent experiments in the mammalian
retina~\cite{Deans02} show that knocking out the gene responsible for
connexin-36 (which accounts for the neuron-neuron gap junction
channels) leads to a dramatic decrease in the dynamical range of
ganglion cell responses.

However, the number of receptor neurons in a given sensory subsystem
is usually limited by several factors, from metabolic costs to sheer
occupied space (for instance, each olfactory glomerulus of the rabbit
receives input from ${\cal O}(10^4)$ sensory
neurons~\cite{Shepherd}). This leads to the second biological
suggestion. Fig.~\ref{fig:dynrangenstate} suggests that for a given
fixed size of the subsystem, it would be advantageous for an organism
to set the refractory period of the receptor neurons to the value
$n_{max}$ that yields the maximum dynamical range. Since $n_{max}(N)$
seems to be an increasing function, we could expect the size of
connected neuron patches and the refractory period of the neurons
(which classically are two independent parameters) to be
correlated. In other words, smaller patches would tend to have neurons
with shorter refractory period. Naturally, it is important to check
the robustness of these results in other models of excitable media
before specific experimental tests could be proposed.

\section{Concluding Remarks}
\label{conclusion}

We have presented results concerning the nonlinear response of the
two-dimensional GHCA to a Poisson stimulus. In particular, the
coupling between the excitable elements has been shown to greatly
enhance the dynamical range due to the collective phenomenon of
self-limited amplification of excitable waves. 

We stress that we do not claim this is the {\em only\/} explanation
for the enhancement of the dynamical range. Variation of the values of
$r_{10}$ and $r_{90}$ within the population of receptor neurons, for
instance, is another hypothesis. In that scenario, different stimulus
ranges would ``recruit'' different groups of neurons, therefore
leading to an overall enhancement of the dynamical range (even if the
neurons were {\it uncoupled\/}). Theoretical work in this direction by
Cleland and Linster~\cite{Cleland99} has shown that this could in
principle be achieved by heterogeneous overexpression of odorant
receptors in the olfactory epithelium. Note, however, that doubling
the dynamical range of the ensemble (as compared to that of individual
neurons) would require the order of a hundred-fold overexpression of
receptors, in that model. As Cleland and Linster point out, ``measured
spare receptor capacities in intracortical and culture systems studied
to date are typically less than twofold''~\cite{Cleland99}.

Interestingly, the different proposed mechanisms are not mutually
exclusive. In fact, ``recruitment'' as described above could in
principle coexist with the self-limited amplification we propose. And
both could cooperate with further nonlinear post-processing of the
signals (e.g. by mitral cell integration in the olfactory bulb). We
believe that the strength of our model lies on two central issues. The
first one is experimental evidence: the model suggests a functional
role for the electrical synapses by gap junctions which have been
experimentally found in the sensory periphery. The second point is
that the mechanism is {\em simple\/}. Indeed, the model relies more
generally on lateral excitation, and could in principle even forego
gap junctions specifically. In the olfactory nerve, for instance, it
could be implemented via the electrical coupling mediated by ephaptic
interactions among neighboring axons, which have recently been modeled
by Bokil et al.~\cite{Bokil01}. Neuronal circuits where mutual
bidirectional excitation by chemical synapses occur could also be
subjected to a similar modeling.

The general question of how brains represent sensory stimuli is a
classical and very difficult one, and has been addressed by physicists
like Fechner, Plateau and Maxwell. Our model addresses only part of
the problem, containing a mechanism to enhance the coding of a very
basic (perhaps the most basic) property of the senses: intensity of
the stimulus.  

Consider, for instance, the retina. Our primary concern is with
electrical coupling not at networks of non-spiking neurons (as cones,
horizontal cells etc) but at excitable networks of spiking cells. So
we are interested in wave generation and synchronization at the first
relevant spiking network (ganglion cell layer, in the case of vision)
as a candidate mechanism for intensity coding in the output of optic
nerve fibers. So, the ``input'' level $r$ presented in our model
should be interpreted as the input to the appropriate excitable
network which we speculate to be situated at the ganglion level, not
the raw input to the retina. There is experimental evidence that the
spikes of ganglion cells are correlated with small latency (typical of
electrical coupling)~\cite{Bair99,Hidaka02}, which is consistent with
our model. For specificity coding, on the other hand, other mechanisms
seem best suited, such as lateral inhibition (which is well
established in the literature, occurring in previous
layers~\cite{Purves}). The olfactory epithelium, on the other hand,
has no known topographical structure in the expression of different
receptors, presenting ``mosaic'' receptor
expression~\cite{Simoes04}. In this case, the excitable elements of
the model could be thought of as unmielynated axons inside an
olfactory nerve fascicle (which are known to pertain to a single class
of receptors and innervate a single glomerulus). Thus the model gives
a potential explanation to the large dynamical range observed in the
intensity coding of individual glomeruli, but not to the much more
difficult problem of how the identity of different smells are
combinatorially coded by the several glomeruli.

Real neurons are obviously much more complex than the our simple CA
model: refractory periods are not necessarily absolute, responses can
come in form of bursts, adaptive firing etc. However, we have also
examined more biologically realistic models previously, leading to
results similar to those of the GHCA~\cite{Copelli02,Copelli05a}. The
robustness of the results presented in this contribution with respect
to different excitable models, lattice connectivity and stimulus
statistics is the subject of ongoing research. For instance,
non-homogeneous two-dimensional lattices have been known to produce
self-sustained activity (spiral waves) since the seminal work of
Wiener and Rosenblueth~\cite{Wiener46}. It remains to be investigated
{\it how\/} to connect excitable elements in order to maximize the
dynamical range of the collective response, but also suppressing
spiral waves.

Despite its lack of biological realism, the CA approach is extremely
useful since it allows the simulation of large networks and even the
calculation of collective properties under appropriate
approximations. Indeed, analytical tools such as hierarchical mean
field approximations~\cite{Tania} are available for studying cellular
automata. We have obtained good results for the one-dimensional system
with a mean field approximation at the pair level~\cite{Furtado},
which analytically reveals the low-$r$ signal compression. The single
site mean field approximation for the GHCA fails in hypercubic
lattices of any dimension, and we are currently investigating the next
hierarchical step to analyze the two-dimensional system. It is
interesting to point out, however, that unless the mean field
calculation contains finite size corrections, it may not be the best
tool for understanding the phenomena occurring in finite cell
assemblies. As has been previously noticed regarding the olfactory
nerve fascicles, many interesting biological phenomena occur precisely
at intermediate values of the lattice size, which emphasizes the
usefulness of computer simulations in addiction to analytical results
that must assume large $N$ (thermodinamic) limits.

\paragraph{Acknowledgments} 
Invaluable discussions with R. F. Oliveira and A. C. Roque are
gratefully acknowledged.  The authors also thank the anonymous
referees for interesting suggestions and comments. MC acknowledges
support from Projeto En\-xo\-val (UFPE), CNPq, FACEPE, CAPES, and
special program PRONEX. OK acknowledges support from FAPESP and CNPq.

\bibliographystyle{unsrt}

\end{document}